\documentclass[aps,preprint,amsmath,amssymb]{revtex4}
\usepackage{graphicx}

\begin{document}

\title
{\Large \bf
 Tunneling Conductance of The Graphene SNS Junction with a Single Localized Defect
}

\author{
Dima Bolmatov\footnote{e-mail: bolmat@phys.nthu.edu.tw} and Chung-Yu Mou}
\affiliation{Department of Physics, National Tsing Hua University, Hsinchu 300, Taiwan \\
 National Center for Theoretical Sciences, Hsinchu 300,Taiwan}


\begin{abstract}
We study the electronic transport in a graphene-based superconductor-normal(graphene)-superconductor (SNS) junction by use of the Dirac-Bogoliubov-de Gennes equation. We consider the properties of tunneling conductance through an undoped strip of graphene with heavily doped superconducting electrodes in the dirty limit $l_{def}\ll L \ll\xi$. We find that spectrum of Andreev bound states  are modified in the presence of  single localized defect in the bulk. The minimum tunneling conductance remains the same and this result doesn't depend on the actual location of the imperfection.

\end{abstract}
\maketitle
               
\section{Introduction}
Graphene, namely a monolayer of graphite, is formed by carbon atoms on a two-dimensional honeycomb lattice. In graphene, due to its unique band structure
whose valence and conductance bands touch at two inequivalent Dirac points (often referred to as $K$ and $K^{'}$) of the Brillouin zone, the electrons around the Fermi
level obey the massless relativistic Dirac equation which
results in the linear energy dispersion relation. Recent exciting developments in transport experiments on graphene has stimulated theoretical studies of superconductivity phenomena in this material, which has been recently fabricated \cite{Nov-1,Zhan-1}. A number of unusual features of superconducting state have been predicted, which are closely related to the Dirac-like spectrum of normal state excitation\cite{Ben-A,Lee-1}.
In particular, the unconventional normal electron dispertion has been shown to result in a nontrivial modification  of Andreev reflection and Andreev bound states in Josephson junctions with supercomducting graphene electrodes\cite{Mai-1,Ash-1,Bol-1}.

Other interesting consequences of the existence of Dirac-like quasiparticles can be understood by studying superconductivity in graphene\cite{Mou-1,Khv-1,Ale-1,Mor-1,Fog-1}. It has been suggested that superconductivity can be induced in graphene layer in the presence of a superconducting electrode near it via proximity effect \cite{Ben-1,Volk-1,Tit-1}.

In this work, we study Josephson effect and find bound state in graphene for tunneling SNS junction with the presence of a single localized defect\cite{Mou-2}. In this study, we shall concentrate on SNS junction with normal region  thickness $L\ll\xi$, where $\xi$ is the superconducting coherence length, and width $W$ which has an applied gate voltage $U$ across the normal region\cite{Mou-3,Sal-1}. In the frame of the dirty limit $l_{def}\ll L \ll\xi$ cosidered by Kulik and Omelyanchuk  we investigate tunneling condactance in SNS junction with presense a single localized defect and find that Andreev levels are modified,  the minimum tunneling conductance remains the same\cite{Kul-1,And-1,Le-1}.

\section{Tunneling conductance of the graphene superconductor/normal/superconductor junction with a single localized defect}
\begin{figure}
  \centering 
 \includegraphics[scale=0.7]{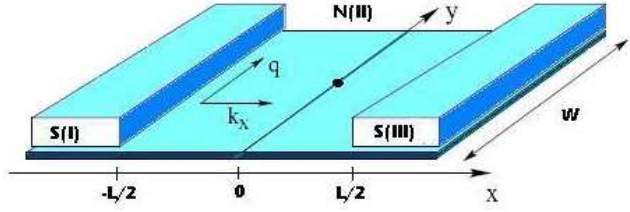} 
	\caption{A graphene undoped ribbon is contacted by two superconducting leads. The charge carriers tunnel from one lead to another via multiple tunneling states formed in the graphene strip. An defect placed inside the strip.}
	\label{fig-1}
	\end{figure}
We consider a SNS junction with a single localized defect which is involved in a graphene sheet of width $W$ lying in the $x-y$ plane  extends from $x=-L/2$ to $x=L/2$ while the superconducting region occupies $|x|>L/2$ (see Fig. 1). The SNS junction can then be described by the Dirac-Bogoliubov-de-Gennes (DBdG) equations\cite{Ben-2},
\[ \left( \begin{array}{c}
\begin{array}{cc}
H_{s}-E_{F}+U & \Delta \\ \Delta^{*} & E_{F}-U-H_{s}
 \end{array}
 \end{array}
 \right)\psi_{s}=\epsilon\psi_{s} \]
Here , $\psi_{s}=(\psi_{As},\psi_{Bs},\psi_{A\overline{s}}^{*},-\psi^{*}_{B\overline{s}})$, $\psi=(u_{1},u_{2},v_{1},v_{2})$ are the $4$ component wave functions for the electron and hole spinors, the index $s$ denote $K$ or $K^{'}$ for electrons or holes near $K$ and $K^{'}$ points, $\overline{s}$ takes values $K(K^{'})$ for $s=K(K^{'})$, $E_{F}$ denotes the Fermi energi, $A$ and $B$ denote the two inequivalent sites in the hexagonal lattice of graphene, and the Hamiltonian $H_{s}$ is given by
\begin{eqnarray}\label{1}
	H_{s}=-i\hbar v_{F}[\sigma_{x}\partial_{x}+sgn(s)\sigma_{y}\partial_{y}]
\end{eqnarray}
In Eq. \ref{1}, $v_{F}$ denotes the Fermi velocity of the quasiparticles in graphene and $sgn(s)$ takes values $\pm$ for $s=K(K^{'})$. The $2\times 2$ Pauli matrices $\sigma_{i}$ act on the sublattice index. The excitation energy $\epsilon>0$ is measured relative to the Fermi level(set at zero). The electrostatic potential $U$ and pair patential $\Delta$ have step function profiles, as in the case of a semiconductor two-dimensional electron gas \cite{Volk-2,Fag-1,Gla-1},
\[ U(x) = \left\{
 \begin{array}{cc}
 &  -U,   \ \mbox{if} \   x<-L/2, \\
 &   0,   \ \mbox{if} \  |x|<L/2, \\
 &  -U,   \  \mbox{if} \  x>L/2.
  \end{array} \right .\]

\[ \Delta(x) = \left\{
 \begin{array}{cc}
  \Delta_{0}\exp(i\phi/2), & \mbox{if} \  x<-L/2, \\
   0                     ,                     & \mbox{if}  \left| x\right|<L/2,  \\
 \Delta_{0}\exp(-i\phi/2), & \mbox{if} \ x>L/2.
  \end{array} \right . \]
The reduction of the order parameter $\Delta(x)$ in the superconducting region on approaching the SN interface is neglected; i.e., we approximate parameter $\Delta(x)$ as we've done it above. As discussed by Likharev \cite{Lik-1}, this approximation is justified if the weak link has length and width much smaller than $\xi$. There is no lattice mismatch at the NS interface, so the honeycomb lattice of graphene is unperturbed at the boundary, the interface is smooth and impurity free.

Solving the DBdG equations, we gain the wave-functions in the superconducting and the normal regions. In region $I(III)$, for the DBdG quasiparticles moving along the $\pm x$ direction with a transverse momentum $k_{y}=q$ and energy $\epsilon$, the wave-functions are given by
 \[ \Psi^{+}=exp(iqy+ik_{s}x+\kappa mx)\left( \begin{array}{c}
\begin{array}{cc}
\exp(-im\beta) \\ \exp(i\gamma-im\beta) \\ \exp(-im\phi/2) \\ \exp(i\gamma-im\phi/2) 
 \end{array}
 \end{array}
 \right) \] 

\[ \Psi^{-}=exp(iqy-ik_{s}x+\kappa mx)\left( \begin{array}{c}
\begin{array}{cc}
\exp(im\beta) \\ \exp(-i\gamma+im\beta) \\ \exp(-im\phi/2) \\ \exp(-i\gamma-im\phi/2) 
 \end{array}
 \end{array}
 \right) \] 
 The parameters $\beta$, $\gamma$, $k_{0}$, $\kappa$ are defined by $\beta=\arccos(\epsilon/\Delta_{0})$, $\gamma=\arcsin[\hbar v_{F} q/(U_{0}+E_{F})]$, $k_{s}=\sqrt{(U_{0}+E_{F})^{2}/(\hbar v_{F})^{2}-q^{2}}$, $\kappa=(U_{0}+E_{F})\Delta_{0}\sin(\beta)/(\hbar^{2} v_{F}^{2} k_{s})$ and $m=\pm$ denotes region $I(III)$, $m=+$ for $I$ and $m=-$ for $III$ correspondently. Further we assumed that the Fermi wave length $\lambda^{'}_{F}$ in the superconducting region much smaller than the wave length $\lambda_{F}$  in the normal region and $U_{0}\gg E_{F},\epsilon$. Since $|q|\leq E_{F}/\hbar v_{F}$, this regime of a heavely doped superconductor corresponds to the limits $\gamma\rightarrow 0$, $k_{s}\rightarrow U_{0}/\hbar v_{F}$, $\kappa\rightarrow (\Delta_{0}/\hbar v_{F})\sin(\beta)$.
 
The region $II$ consists of three areas: $a,b,c$ (see Fig. 2). We solve the DBdG equations for the area $b$, while area $c$ is the area where placed defect and area $a$ would be extended and matched with superconducting regions. The two valleys $s\pm$ decouple, and we can solve equations separately for each valley, $H_{s}\psi_{s}=(\epsilon+sE_{F})\psi_{s}$, $H_{s}=H_{0}+sV(r)\sigma_{z}$. The term proportional to $\sigma_{z}$ in Hamiltonian is a mass term confining the Dirac electrons in the area $b$. Rewrite the Hamiltonian in the cylindrical coordinates and since $H_{s}$ commutes with $J_{z}=l_{z}+\frac{1}{2}\sigma_{z}$, its electron-eigenspinors  $\psi_{e}$  are eigenstates of $J_{z}$ \cite{Ben-ring},

\begin{figure}
	\centering
  \includegraphics[scale=0.9]{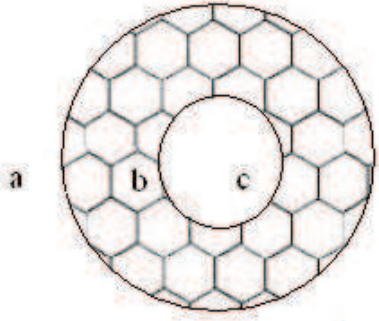}
	\caption{The normal region $II$ is devided by three areas $a,b,c$. The DBdG equations are solved for area $b$ and determined by the "infinite mass" boundary conditions induced by $V(r)\rightarrow +\infty$  in $a$ and $c$ areas correspondently.}
	\label{fig:Ring}
	\end{figure} 
	
 \[ \Psi_{e}(r,\alpha)=\left( \begin{array}{c}
\begin{array}{cc}
\exp(id(n-1/2)\alpha) J_{d(n-1/2)}(k(\epsilon)r) \\ \exp(id(n+1/2)\alpha) J_{d(n+1/2)}(k(\epsilon)r)
 \end{array}
 \end{array}
 \right) \] 
with eigenvalues $n$, where $n$ is a half-odd integer, $n=d\frac{1}{2},d\frac{3}{2}, \ldots$ and $J_{d(n-1/2)}(k(\epsilon)r)$ is the Bessel function of $(n-1/2)$ order. In the $x-y$ plane $d$ denotes moving direction of correspondent quasiparticle, $d=+$ for the quasiparticle moving toward  $x=L/2$ and $d=-$ for the quasiparticle moving toward $x=-L/2$ direction correspondentely. Further we are interested in to find zero energy states\cite{Tit-1}. In this case the DBdG equations posses a general symmetry with respect to the change in the sign of energy 
\begin{eqnarray}
	\epsilon\rightarrow-\epsilon, \ \ i\widehat{\sigma}_{y}\widehat{u}^{*}\rightarrow\widehat{v}, \ \ i\widehat{\sigma}_{y}\widehat{v}^{*}\rightarrow-\widehat{u},
\end{eqnarray}
where we denote $\widehat{u}=(u_{1},u_{2})$ and $\widehat{v}=(v_{1},v_{2})$. Thus, for a set of zero modes ($\widehat{u}_{i}$,$\widehat{v}_{i}$) enumerated by a certain index $i$ we should have,
\begin{eqnarray}
	  \widehat{v}_{i}=i\widehat{\sigma}_{y}\widehat{u}^{*}_{j}, \ \widehat{u}_{i}=-i\widehat{\sigma}_{y}\widehat{v}^{*}_{j}.
\end{eqnarray}
In the same manner as for electron hole-spinors have form
\[ \Psi_{h}(r,\alpha)=\left( \begin{array}{c}
\begin{array}{cc}
-\exp(id(n+1/2)\alpha^{'}) J_{d(n+1/2)}(k^{'}(\epsilon)r) \\ \exp(id(n-1/2)\alpha^{'}) J_{d(n-1/2)}(k^{'}(\epsilon)r)
 \end{array}
 \end{array}
 \right) \] 	
with the definitions 
\begin{eqnarray}
	\alpha(\epsilon)=\arcsin[\hbar v_{F} q/(\epsilon+E_{F})], \ 	\alpha^{'}(\epsilon)=\arcsin[\hbar v_{F} q/(\epsilon-E_{F})],
\end{eqnarray}
\begin{eqnarray}	k(\epsilon)=(\hbar v_{F})^{-1}(\epsilon+E_{F})\cos(\alpha),\	k^{'}(\epsilon)=(\hbar v_{F})^{-1}(\epsilon-E_{F})\cos(\alpha),
\end{eqnarray}

\begin{figure}
 \centering
 \includegraphics[scale=0.8]{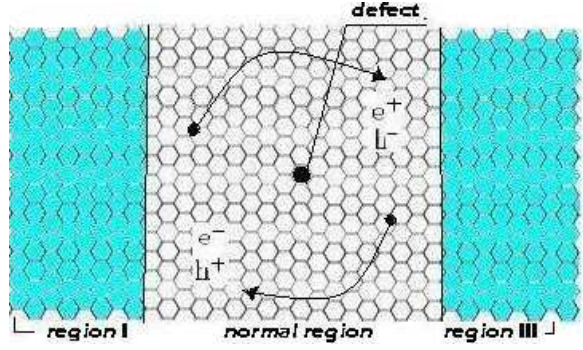} 
 	\caption{Scheme of graphene SNS junction with a single impurity which is placed inside the strip of the region $II$.}
 	\label{fig-3}
  \end{figure} 

The angle $\alpha\in(-\pi/2,\pi/2)$ is the angle of incidence of the electron (having longitudinal wave vector $k$), and $\alpha^{'}$ is the reflection angle of the hole (having longitudinal wave vector $k^{'}$) \cite{Asa-1,Bhat-1}.
To obtain an analytical approximation of the spectrum, we use the asymptotic form of the Bessel functions for large $r$. This indeed  is the desired limit as $rk(\epsilon)\approx r_{def}k(\epsilon)\propto r_{def}/L\ll1$, where $r_{def}$ is the defect radius (the radius of the area $c$) and determine for all eigenvalues $n=d\frac{1}{2}$. In this limit we impose the $"$infinite mass$"$ boundary conditions at $y=0,W$, for which $q_{n}=(n+1/2)\pi/W$ in the area $b$ with $V(r)\rightarrow +\infty$ in the $a$ and $c$ areas consequently . Half-odd integer values $n$ reflect the $\pi$ Berry's phase of closed size of a single localized defect in graphene.

To obtain the subgap ($\epsilon<\Delta_{0}$) Andreev bound states, we now impose the boundary conditions at the graphene. The wave-functions in the superconducting and normal regions can be constructed as
\begin{eqnarray}
&&	\Psi_{I}=a_{1}\psi_{I}^{+}+b_{1}\psi_{I}^{-}, \ \Psi_{III}=a_{2}\psi_{I}^{+}+b_{2}\psi_{I}^{-}, \\
	&& \Psi_{II}=a\psi_{II}^{e+}+b\psi_{II}^{e-}+c\psi_{II}^{h+}+d\psi_{II}^{h-}.
\end{eqnarray}
where $a_{1}(b_{1}),a_{2}(b_{2})$ are the amplitudes of right and left moving DBdG quasiparticles in region $I(III)$ and $a(b)$ and $c(d)$ are the amplitudes of right(left) moving electrons and holes, respectively, in the normal region\cite{Mai-1}.  These wave functions must satisfy the boundary conditions,
\begin{eqnarray}
	\Psi_{I}|_{x=-L/2}=\Psi_{II}|_{x=-L/2}, \ \Psi_{II}|_{x=L/2}=\Psi_{III}|_{x=L/2}.
\end{eqnarray}
Since the wave vector $k_{y}$ parallel to the $NS$ interface and different wave vectors in the $y$-direction are not coupled, we may solve the problem for a given $k_{y}=q$ and  we can consider each transverse mode separately. To leading order in the small parameter $\Delta_{0}L/\hbar v_{F}$ we may substitute $\alpha(\alpha^{'})\rightarrow\alpha(0)$, $k(\epsilon)(k^{'}(\epsilon))\rightarrow k(0)$. After some algebra we obtain equation
\begin{eqnarray}\label{2}
 \cos(2\beta)\left\{ \frac{\sin^{2}(kL)-\cos^{2}(kL)\sin^{2}(\alpha)}{\cos^{2}(kL)\cos^{2}(\alpha)-1}\right\}-\sin(2\beta)\left\{ \frac{\cos(kL)\sin(kL)\sin(\alpha)}{\cos^{2}(kL)\cos^{2}(\alpha)-1}\right\}=\cos(\phi)
\end{eqnarray} 
Eq. \ref{2} differs from equation  obtained for SNS junction without a single defect. Eliminating second term in Eq. \ref{2} we could immediately yield reduction of the equation and it earns a essential form for weak SNS junctions\cite{Tit-1}. The solution  of  Eq. \ref{2} is a single bound state per mode,
\begin{eqnarray}
\epsilon_{n}=\Delta_{0}\sqrt{\frac{2}{A^{2}+B^{2}}(-2CA^{2}-B^{2}+ B\sqrt{1-4CA^{2}(C+1)})}
\end{eqnarray}
where 
\begin{eqnarray}
A=\frac{\sin^{2}(k_{n}L)-\cos^{2}(k_{n}L)\sin^{2}(\alpha)}{\cos^{2}(k_{n}L)\cos^{2}(\alpha)-1},\ B= \frac{\cos(k_{n}L)\sin(k_{n}L)\sin(\alpha)}{\cos^{2}(k_{n}L)\cos^{2}(\alpha)-1}
\end{eqnarray} 
\begin{eqnarray}\label{Trans}
C=\frac{1}{2}+\frac{1}{\tau_{n}}(\frac{1}{2}-\sin^{2}(\frac{\phi}{2})), \ \tau_{n}=\frac{\cos^{2}(k_{n}L)\cos^{2}(\alpha)-1}{\cos^{2}(k_{n}L)\cos^{2}(\alpha)-\cos(2k_{n}L)},
\end{eqnarray}

\begin{figure}
	\centering
	\includegraphics[scale=0.9]{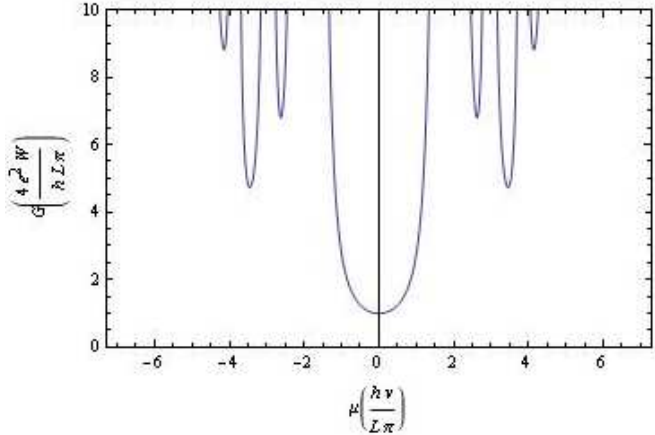} 
	\caption{Tunneling conductance of the graphene SNS junctions with a single localized defect versus Fermi energy, calculated from Eq. \ref{Trans}. The tunneling conductance exhibits oscillatory behavior.}
	\label{fig-4}
\end{figure}

We don't have a simple analytic expression for the $\phi$-dependance but we obtained modified Andreev levels with the presense a single localized defect in the bulk. The conductance of the graphene strip is expressed through the transmission probability by the Landauer formula,
\begin{eqnarray}\label{Con}
	G=g_{0}\sum_{n=0}^{n(\mu)}\tau_{n}, \ \ g_{0}=4e^{2}/h,
\end{eqnarray}
where $n(\mu)\gg1$ is given by $n(\mu)$=Int$(k_{n}W/\pi+1/2)$. Substitution transmission probability into Eq. \ref{Con} gives the conductance versus Fermi energy  (see in Fig. 3,Fig. 4). The result for the minimal conductivity agrees with other calculations\cite{Lud,Zi,Per}, which start from an un-
bounded disordered system and then take the limit of
infinite mean free path $l$. There is no geometry dependence if the limits are taken in that order.

\section{Summary}
In the conclusion, we have shown that tunneling conductance of the graphene SNS junction with a single localized defect has a nonzero minimal value if the Fermi level is tuned to the point of zero carrier concentration.We have demonstrated that the tunneling conductance exhibits oscillatory behavior. Andreev levels are modified, the minimum tunneling conductance remains the same and this result doesn't depend on the actual location of the imperfection. The consideration of  tunneling conductance of SNS junctions with multiple defects is the subject of forthcoming article.

\section{Acknowledgments}
Authors are very indebted to Prof. H.H. Lin for stimulating discussions and thank anonymous referees for valuable and credible comments. We acknowledge support from the National Center for Theoretical Sciences in Taiwan.

\end{document}